\documentclass[prl,aps,twocolumn,floatfix]{revtex4}
\usepackage{amsmath,graphics,epsfig,color,verbatim,ulem}

\renewcommand{\vr}{{\mathbf{r}}} \newcommand{\vk}{{\mathbf{k}}}

\begin{document}
\title{Self consistent GW determination of the interaction strength:
application to the iron arsenide superconductors}

\author{A. Kutepov$^1$, K. Haule$^1$, S.Y. Savrasov$^2$, G. Kotliar$^1$}
\affiliation{$^1$ Department of Physics, Rutgers University,
Piscataway, NJ 08854, USA}
\affiliation{$^2$ Department of Physics, University of California, Davis, CA 95616}


\begin{abstract}
We introduce a first principles approach to determine the
strength of the electronic correlations based on  the fully self
consistent GW approximation.  The approach provides a seamless
interface with 
dynamical mean field theory, and gives good results for well
studied correlated materials such as NiO. Applied to the recently
discovered iron arsenide materials, it accounts for the noticeable
correlation features observed in optics and photoemission while
explaining  the absence of visible satellites in X-ray absorption
experiments and other high energy spectroscopies.
\end{abstract}
\pacs{71.27.+a,71.30.+h} \date{\today}
\maketitle


Many metals, semiconductors, and insulators are well described by
the "standard model" of solid state physics. In this picture the
excitations are band electrons, and their dispersion can be
computed quantitatively  in perturbation theory starting from the
density functional theory using the GW method \cite{Hedin}.  When
this standard model fails, we talk about strongly correlated
electron systems.
The presence of strong correlations is debated with each new
material discovery, as for example in the context of the iron
pnictide superconductors. On the experimental side, controversies
arose because
optical experiments revealed significant mass renormalizations
\cite{Basov,optics-best,optics1} while Xray absorption, core level
spectroscopies and resonant inelastic Xray scattering indicated
the absence of satellite peaks \cite{X-ray1,X-ray2}, which are
standard fingerprints of strong correlations.  Photoemission
studies indicate
that the overall bandwidth is narrowed by a factor of two
\cite{ARPES1,ARPES2} but substantially larger mass
renormalizations are present near the Fermi level
\cite{FrenchExp}.
Similar controversies arose within the first principles
approaches to the treatment of correlations with some theoretical
studies supporting the notion of weak correlations
\cite{Aichorn,Anisimov1,Anisimov2,Mazin,Singh}, while others
advocate a more correlated picture
\cite{ourFeAs,QSi,Craco,Liebsch}. To make progress on this issues
one needs   to develop fully ab initio tools for addressing the
problem of determining the  strength of correlations and test
their predictions  against experiments.

In this letter we introduce a new first principles methodology for
evaluating the strength of the correlations based on the
self-consistent GW method. This approach  has been shown  to
predict accurate total energy~\cite{VonBarth,Kutepov}, and we
expect to obtain reliable estimates for the interaction strength
since this quantity can be thought as  a   second derivative of
the total energy with respect to the occupation of the correlated
orbitals. We test successfully the method on the well studied
example of a correlated material NiO, and then we apply it to a
prototypical iron pnictide BaFe$_2$As$_2$. We  find that the
correlations in iron pnictides are strong, as pointed out in
Refs.~\cite{ourFeAs,QSi,Craco,Liebsch} but unlike earlier studies
our  ab-initio method accounts for the absence of well defined
Hubbard bands in the spectral functions. Our results are thus  in
excellent agreement with experiment and reconcile the results of
apparently conflicting spectroscopies.

%


%
We start with the one-particle electron Green's function in the
solid, $G$,  
which is measurable in photoemission experiments. We split it
into $G^{-1}={G^0}^{-1}+\Sigma$, where ${G^0}$ describes the
non-interacting system of electrons, and $\Sigma$ is the
frequency dependent self-energy. Both $G$ and $\Sigma$ are
matrices in $\vr,\vr'$.
The electrons interact among themselves via the Coulomb
interaction $V_c(\vr,\vr')=\frac{1}{|\vr-\vr'|}$, however, the mobile
electrons screen it and is therefore useful to reformulate the
problem in terms of a screened Coulomb interaction $W$ defined by
$ W = V_c /(1 + V_c \Pi) $ \cite{Ahmblada,Hedin} where $\Pi $ is
the exact polarization function.

Dynamical Mean Field Theory (DMFT) maps the many body problem in the
solid to that of a correlated {\it atomic } shell embedded in an
effective medium.  The medium is described by an energy dependent
Weiss field ${\cal G}^0$, which obeys the following equation
\begin{equation}
  {{\cal G}^0}^{-1}= {G_{local}}^{-1} + \Sigma_{local}.
\label{G0eq}
\end{equation}
Here $G_{local}$ and $\Sigma_{local}$ are the local Green's function
and the local self-energy, respectively.

The electrons in the renormalized atom feel an effective retarded
Coulomb interaction $U(\omega)$. Just like the Weiss field ${{\cal
    G}^0}$ of the atom reflects the delocalizing effect of the medium
at the single particle level, the Weiss field $U(\omega)$ captures the
screening of the interaction due to the presence of the other atoms.
The Weiss field at the two particle level $U(\omega)$ obeys the
following relation
\begin{equation}
U^{-1}= W_{local}^{-1}+ \Pi_{local},
\label{Ueq}
\end{equation}
where $\Pi_{local} $ and $W_{local}$ are the local polarization
function and the local screened interaction, respectively.
The bare local propagators ${{\cal G}^0}(\omega)$ and bare
interaction $U(\omega)$ are chosen so as to give the exact
$G_{local}$ and $W_{local}$ when all the local Feynman
diagrams are summed up.
%
%
%
Eqs.~(\ref{G0eq}) and (\ref{Ueq}) are a version of the extended-DMFT
equations studied for simplified models in Refs.~\cite{Si,Chitra}.
The key idea of this work is to use this approach to estimate the
correlation strength in the solid, and illustrate the power of the
method by a practical realistic self-consistent implementation.

The approach shares ideas with other methods to compute the local
interaction matrix $U$. Like constrained LDA, it defines
correlations on a correlated \textit{orbital}. It adopts the
philosophy of the constrained RPA method \cite{CRPA,Miyake1},
which divides the \textit{bands} into a set that belongs to the
low energy model, and the rest of the bands, which contribute to
screening.  However, instead of the bands, our method uses
orbitals to divide the polarization operator of the lattice  into
a local part, involving the correlated orbital, and the rest, which
screens the local interaction.




We now describe  the steps required for the  practical
implementation of the method and its interface with LDA+DMFT
%
%
\cite{review}:
i) We perform a
{\it fully self consistent} GW calculation \cite{Hedin}.
%
ii) We evaluate $G_{loc}$ and $W_{loc}$ using
the projector
$P(\vr\vr',t L_1 L_2)$, defined in Ref.~\onlinecite{ourDMFT}:
${G_{local}}^t_{L_1 L_2} =  \int d\vr P(\vr\vr',t L_1 L_2) G(\vr \vr')$
and
${W_{local}}^t_{L_4 L_1; L_3 L_2} = \int P(\vr\vr,t L_4 L_1)
W(\vr\vr')P(\vr'\vr',t L_3 L_2)d\vr d\vr' .$
$t$ is the atom index and $L=(l,m)$ is the angular momentum index.
iii) We evaluate
$\Sigma_{loc}(\tau)=W_{local}(\tau)G_{local}(-\tau)$ and
$\Pi_{loc}(\tau) = G_{local}(\tau)G_{local}(-\tau)$. iii) We use
Eq.~(\ref{Ueq}) to evaluate $U(\omega)$, which we now denote by $U^{GW}$.
iv) We also evaluate the hybridization function $\Delta_{GW}(\omega)$
using Eq.~(\ref{G0eq}) and identity ${{\cal G}^0_{GW}}^{-1}=\omega-E_{imp}-\Delta_{GW}$.
%
%
$\Delta_{GW}$ contains the coupling of the correlated orbitals to the
valence states of the system $\Delta_L(\omega)$, and to the semicore
states $\Delta_H(\omega)$, and can thus be represented as
$\Delta(i\omega) = \int d\varepsilon{[\Delta_L(\varepsilon) +
    \Delta_H(\varepsilon)]}/{(i\omega-\varepsilon)}$.
In LDA+DMFT the hybridization to these semicore
state is eliminated resulting in $\Delta_{L}$, which is connected to
GW hybridization by $\Delta_{GW}(i\omega) = \Delta_{L}(i\omega) -
i\omega \alpha$ where $\alpha \approx \int d\varepsilon
\Delta_H(\varepsilon)/\varepsilon^2$.  This factor is then absorbed by
rescaling of the field $\psi\rightarrow\psi/\sqrt{1+\alpha}$ and
consequently the interaction matrix used in the LDA+DMFT calculation
becomes $U_{LDA+DMFT} = U^{GW}(\omega=0)/(1+\alpha)^2$. This
renormalization is usually very small, and in BaFe$_2$As$_2$ is
$\alpha\approx 0.05$.

We use this fully ab initio method to determine the interaction
matrix strength $U^{GW}$ and the occupancy of the $d$ orbital
$n_d$, which fixes the double-counting correction of LDA+DMFT.
With this input, the LDA+DMFT method becomes a fully ab-initio
method.


%

The effective interaction obtained with this method is a general
symmetric tensor with four indices
$\sum_{\{m_i\},\sigma\sigma'}U_{m_4,m_3,m_2,m_1}\psi^\dagger_{m_4\sigma}\psi^\dagger_{m_3\sigma'}\psi_{m_2\sigma'}\psi_{m_1\sigma}$.
It is useful to inquire to which extent this interaction can be
approximated in terms of Slater integrals $F^{\{l\}}_k$, where $k$
runs over $0,...2l$.  The optimal determination of this parameters is
done with the projector
\begin{eqnarray}
F_k^{\{l\}} = \sum_{m_1,m_2,m_3,m_4}\frac{1}
{{\cal N}_{l,k}}\frac{4\pi}{2k+1}
\langle Y_{l m_4}|Y_{k\; m_4-m_1}|Y_{l m_1}\rangle\nonumber\\
\times U^{GW}_{m_4 m_3 m_2 m_1}
\langle Y_{l m_3}|Y^*_{k\; m_2-m_3}|Y_{l m_2}\rangle
\end{eqnarray}
Here ${\cal N}_{l,0}=(2l+1)^2$, ${\cal N}_{l=2,k=1}=5(2/7)^2$ and
${\cal N}_{l=2,k=2}=(10/21)^2$.
The quality of the projection is excellent and can be seen by
 recomputing  the Coulomb repulsion from  the Slater integrals
%
and comparing the resulting $U^{atom}$ with the full $U$ matrix.
%
%
We mention in passing that the naive Hartree-Fock like estimation of
Slater integrals 
$J=\langle U_{m m' m m'}\rangle_{m\ne m'}$, $F_2=14/1.625\, J$ and $F_4 =
8.75/1.625\, J$, can lead to a substantial underestimation of Slater
integrals.


We first test our method in an arc-typical charge transfer insulator
NiO. We get the following static values of the Slater integrals
$F_0=7.9\,$eV, $F_2=10\,$eV, and $F_4=6.7\,$eV. If the Hund's
parameter $J$ is computed from $F_2$ ( $F_4$ ) we get
$J(F_2)=1.16\,$eV ( $J(F_4)=1.24\,$eV ). When these parameter are used
in LDA+DMFT, the agreement between the theory and experiment is very
good \cite{Kunes-NiO}. We mention in passing that when the GW screened
interaction and polarization are computed from the LDA  Kohn-Sham
states (non self-consistent GW) we get slightly smaller interaction
strength $F_0\approx 7.2\,$eV.



\begin{figure}[!ht]
\centering{
  \includegraphics[width=1.0\linewidth]{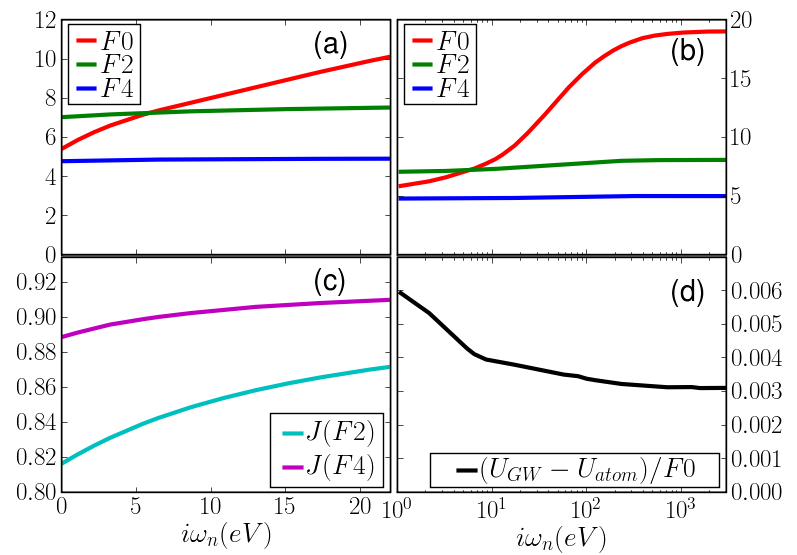}
  }
\caption{ (Color online) (a),(b) Slater integrals versus Matsubara frequency
  as computed by fully self-consistent GW method,
  (c) corresponding Hund's coupling strength $J$, and (d) the difference
  between the GW Coulomb interaction and its Slater parameterization.}
\label{U_B122}
\end{figure}
Next we turn to the Coulomb repulsion in
BaFe$_2$As$_2$. Fig.~\ref{U_B122}(a) and (b) show the frequency
dependence of the Slater integrals for Fe-$3d$ orbitals on imaginary
frequency axis in linear and log scale, respectively.
At very high frequency, the interaction is unscreened and approaches
its atomic value. The density-density Coulomb interaction $F_0$ is
strongly screened in the solid, while the higher multipoles $F_2$ and
$F_4$ are much less energy dependent, and almost equal in solid as in
the atom.

The static Coulomb interaction $F_0$ is estimated to be no less then
$5\,$eV, larger then previously estimated by constrained
LDA~\cite{Anisimov1} and constrained RPA~\cite{Aichorn}. We want to
remark that the self-consistency of GW is important in this material,
because the non-self consistent version of GW leads to weaker
interaction strength $F_0\approx 3.4\,$eV.


The higher order multipoles $F_2$ and $F_4$ show only a weak frequency
dependence. The highest multipole $F_4$ is less screened then $F_2$,
and hence a single number $J$ does not parameterize the form of the
Hund's coupling very well, as $J(F_2)\ne J(F_4)$ in Fig.~\ref{U_B122}(c).
Finally Fig.~\ref{U_B122}(d) shows that the Slater parametrization of
the GW Coulomb interaction is remarkably accurate in
BaFe$_2$As$_2$, with error less then 6\%.

We now turn to the spectral properties of the BaFe$_2$As$_2$.  Earlier
5-band model LDA+DMFT calculations \cite{ourFeAs} displayed important
mass renormalization at low energies ($m^*/m\approx 3-5$), but also
showed a sharp lower Hubbard band.
We now perform computations by newly implemented charge self-consistent
LDA+DMFT(CTQMC) method, based on WIEN2K~\cite{Wien2K}, and explained in
detail in Ref.~\onlinecite{ourDMFT}.
%
%
%
The GW estimate for the Slater integrals, renormalized by
$1/(1+\alpha)^2=0.91$ due to elimination of the hybridization with
semicore states, are $F_0=4.9\,$eV, $F_2=6.4\,$eV and $F_4=4.3\,$eV. We
use the standard localized limit double-counting, which gives for the
valence $n_d=6.2\pm0.05$, in perfect agreement with GW estimate
$n_d=6.2$, hence there is no uncertainty in appropriateness of the
chosen double-counting correction. The quasiparticle mass
renormalization obtained by DMFT is $Z\sim 1/2$ for lighter $x^2-y^2$
and $z^2$ orbitals and $Z\sim 1/3$ for heavier $xz,yz$ and $zx$
orbitals.

\begin{figure}[!ht]
\centering{
  \includegraphics[width=0.9\linewidth]{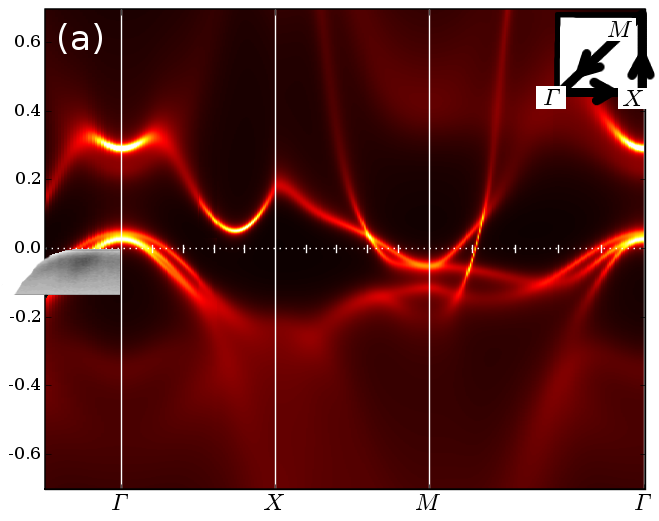}
  \includegraphics[width=0.9\linewidth]{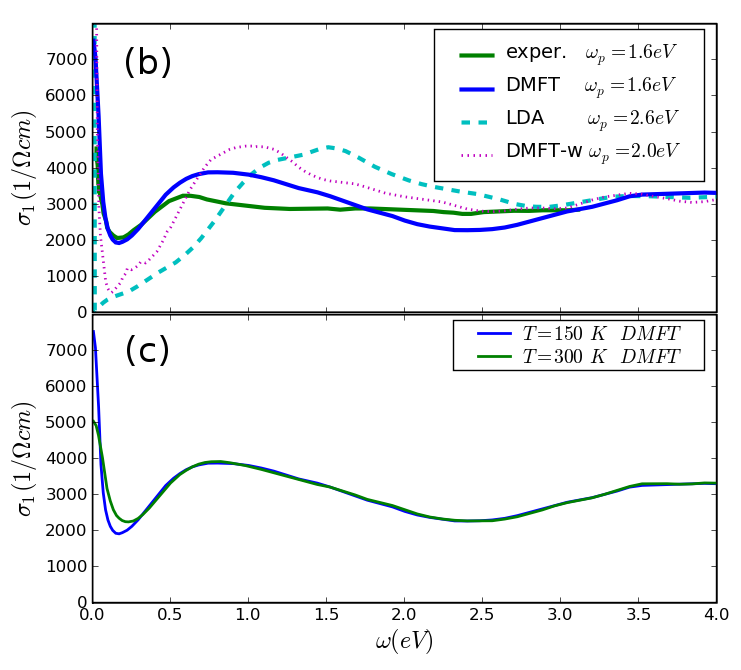}
  }
\caption{ (Color online) (a) $A(\vk,\omega)$ of BaFe$_2$As$_2$ at $T=150\,$K as
  computed by LDA+DMFT. The inset shows the path in momentum space,
  while the grey inset shows the ARPES intensity from
  Ref.~\cite{FrenchExp}. (b) Optical conductivity of LDA+DMFT method
  (DMFT) and its comparison with experiment of
  Ref.~\onlinecite{optics1} (exper.). Also shown is the LDA optical
  conductivity and LDA+DMFT conductivity with substantially smaller
  $F_0$ of Ref.~\cite{Aichorn} (DMFT-w). The legend contains the
  strength of the Drude peak, which is broadened only due to
  electron-electron interactions.  (c) Temperature dependence of the
  optical conductivity within LDA+DMFT.  }
\label{Aopt}
\end{figure}
In Fig.~\ref{Aopt}(a) we plot the spectral function $A(\vk,\omega)$
along the path shown in the inset.  There are three circular hole
pockets at $\Gamma$ and two electron pockets at $M$, in agreement with
ARPES \cite{Hding,Xu} and LDA. The two smaller pockets are degenerate
and their crossing occurs at $0.166\pi/a$ and $0.28\pi/a$, in good
agreement with ARPES \cite{FrenchExp}, where the pocket size was
estimated to $0.14\,\pi/a$ and $0.28\,\pi/a$, respectively.
The Fermi velocities at
$\Gamma$ point towards $X$, predicted by our method, are $\sim 0.45\,
\textrm{eV\AA}$, more then twice smaller then in LDA. The velocity is
in reasonable agreement with experiment where somewhat different
velocities for the two pockets are estimated to be
$0.43\,\textrm{eV\AA}$ and $0.32\,\textrm{eV\AA}$ \cite{FrenchExp}. We also
overlay the ARPES intensity from Ref.~\onlinecite{FrenchExp} on our spectral
dispersion to emphasize good agreement.

The optical conductivity is also a strong test of the correlation
strength, as pointed out in Ref.~\onlinecite{Basov}.
%
%
The strong reduction of Drude weight and the presence of mid-infrared
peak at $\sim 0.6\,$eV was noticed in
Refs.~\onlinecite{optics1,optics-best}. In Fig.~\ref{Aopt}(b) we show
optics obtained by LDA and by DMFT, and we compare it to experimental
results of Refs.~\onlinecite{optics1}. Although LDA gives a reasonable
order of magnitude for optics, it clearly disagrees with experiments
in strength of the Drude peak ($\omega_p\approx 2.6\,$eV) and position
of the mid-infrared peak, coming from the interband transitions. In
contrary our DMFT results give smaller Drude peak of strength
$\omega_p=1.6\,$eV, in very favorable agreement with
experiments~\cite{optics1,optics-best}. We also notice similar width
of the Drude peak in DMFT and experiments, which shows that the most
important channel for scattering in this material is the
electron-electron scattering. Finally, the position of the
mid-infrared peak, which LDA predicts at frequency $\sim 1.2\,$eV,
appears around $0.6\,$eV in DMFT, in very favorable agreement with
experiment \cite{optics1,optics-best}.

To demonstrate the sensitivity of the optical conductivity to the
strength of the correlations, we carried out LDA+DMFT calculation
using parameters of Ref.~\onlinecite{Aichorn}, $F_0=2.69\,$eV,
$J=0.79\,$eV, and $n_d=6.53$. The resulting quasiparticle
renormalization amplitude is $Z \sim 0.6$, in very good agreement with
results of Ref.~\onlinecite{Aichorn}. In Fig.~\ref{Aopt}(b) we show
optical conductivity thus obtained with label DMFT-w. We noticed that
neither Drude peak weight ($\omega_p = 2.0\,$eV) nor the position of
the mid-infrared peak ($\approx 1\,$eV) is in good agreement with
experiments, thus confirming that BaFe$_2$As$_2$ should not be
regarded as weakly correlated material.

Finally, Fig.~\ref{Aopt}(c)
shows the temperature dependence of optical conductivity as obtained
by DMFT. While the interband transitions are roughly temperature
independent, the Drude peak width and strength is temperature
dependent, substantially sharpening at $150\,$K compared to $300\,$K,
which is the consequence of the coherence incoherence crossover in
this temperature range, discussed in Ref.~\cite{ourNewJournal}.

\begin{figure}[!ht]
\centering{
  \includegraphics[width=0.9\linewidth]{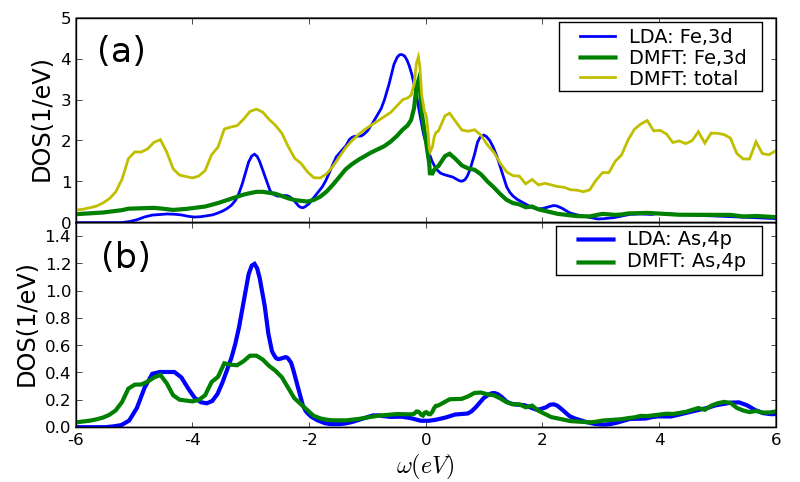}
  }
\caption{ (Color online) Total and partial density of states of LDA+DMFT method
  compared to LDA density of states.}
\label{DOS-122}
\end{figure}
Fig.~\ref{DOS-122}(a) show the total density of states (DOS) and
Fe-$3d$ partial DOS. Panel (b) shows the As-$4p$ partial DOS.
Comparing LDA+DMFT partial DOS to LDA partial DOS, we notice that
apart from the renormalization of the low energy quasi-particles,
and broadening of the high energy features, there is only little
difference between LDA and LDA+DMFT momentum averaged spectral
functions. This is in agreement with X-ray absorption spectroscopy
\cite{X-ray2}, where good agreement between LDA and the
experiments was pointed out.
Given the strong correlation effects present in optics and low
energy ARPES, it is unusual that no clear Hubbard-like satellites
of the atomic like $3 d^5$ state can be identified in local
density of states.
%

The  DMFT valence histogram \cite{Nature}, describing the
probability of finding each Fe-$3d$ atomic  configuration   in
the solid as a function of the renormalized energy of the atomic
state  sheds light on the unusual metallic state of the iron
pnictides. In a weakly correlated metal, almost all the atomic
configurations   are significantly  present in the ground state
of the solid and their energy  vary over the scale of the
hybridization which represents the bandwidth of the metal. In
correlated oxides, on the other hand, only a few atomic states in
each valence have substantial weight,  which results in sharp
Hubbard bands.
%
%
In BaFe$_2$As$_2$, the probability of the atomic ground state
with valence $N=6$, $N=7$ and $N=5$ is only 0.014, 0.01, and
0.007, respectively. Other states have smaller probability, but
remarkably {\it all} atomic states with valence 5, 6, and 7 have
finite probability larger then 0.001.  The large occupancy of the
extremely large number of atomic configurations is reminiscent of
an itinerant system. On the other hand, unlike the weakly
correlated situation,  the   spread of the multiplets    of the
$N=5$ states, (coming from the Slater integrals $F_2$ and $F_4$)
is $\sim 7\,$eV  similarly the atomic states with $N=6$ span an
energy range of $6.5\,$eV. This scale, represents the width of the
Hubbard bands and is  very large, much larger than the scale of
the hybridization ($2.\,$eV).


%
%
We stress that the absence of clear atomic-like satellite excitations
is not due to weak correlations in FeAs materials, as suggested in
Refs.~\onlinecite{Anisimov1,Anisimov2,Aichorn}, but rather due to the
strength of the atomic multiplet splittings and due to the broad
bandwidth of the highly polarizable As states.  This situation, arises
for the parameters determined from the self consistent GW method. It
is significantly different from what is found in the oxides, and is
captured by the charge self-consistent LDA+DMFT calculation.

\section{Acknowledgement}
We are grateful to N. Zein, A. Georges and A. Aichorn for discussion.
K.H was supported by Grant NSF DMR-0746395 and Alfred P. Sloan
foundation. G.K. and A.K. were suported by NSF DMR-0906943. KH and GK
acknowledge support of NSF DMR 0806937.  S.S. was supported by DOE
SciDAC Grant No. SE-FC02-06ER25793.

\end{document}